\documentclass[letterpaper,12pt]{article}

\usepackage{setspace}
\usepackage{dgjournal}          
\usepackage{color,soul}
\usepackage{amssymb,amsmath}
\usepackage{multirow}
\usepackage{graphics}
\usepackage[authoryear,comma,longnamesfirst,sectionbib]{natbib} 
\usepackage{lscape}
\newcommand{\hlc}[2][yellow]{{\sethlcolor{#1}\hl{#2}}}
\usepackage{booktabs}
\begin{document}



\begin{center} Building an NCAA men's basketball predictive model and quantifying its success \end{center}

\doublespacing
\section{Introduction}

Each March, more than an estimated 50 million Americans fill out a bracket for the National Collegiate Athletic Association (NCAA) men's Division 1 basketball tournament \citep{Atlantic}. While paid entry into tournament pools is technically outlawed, prosecution has proved rare and ineffective; an estimated \$2.5 billion was illegally wagered on the tournament in 2012 \citep{LAT, BusinessWeek}. 

Free tournament pools are legal, however, and Kaggle, a website that organizes free analytics and modeling contests, hosted its first college basketball competition in the early months of 2014. Dubbed the `March Machine Learning Mania' contest, and henceforth simply referred to as the Kaggle contest, the competition drew more than 400 submissions, each competing for a grand prize of \$15,000, which was sponsored by Intel. We submitted two entries, detailed in Section \ref{MC}, one of which earned first place in this year's contest. 

This manuscript both describes our novel predictive models and quantifies the possible benefits, with respect to contest standings, of having a strong model. First, we describe our submission, building on themes first suggested by \cite{carlin1996improved} by merging information from the Las Vegas point spread with team-based possession metrics. The success of our entry reinforces longstanding themes of predictive modeling, including the benefits of combining multiple predictive tools and the importance of using the best possible data. 

Next, we use simulations to estimate the fraction of our success which can be attributed to chance and to skill, using different underlying sets of probabilities for each conceivable 2014 tournament game. If one of our two submissions contained the exact win probabilities, we estimate that submission increased our chances of winning by about a factor of 50, relative to if the contest winner were to have been randomly chosen. Despite this advantage, due to the contest's popularity, that submission would have had no more than about a 50-50 chance of finishing in the top 10, even under the most optimal of conditions.
 
This paper is laid out as follows. Section 2 describes the data, methods, and scoring systems pertinent to predicting college basketball outcomes. Section 3 details our submission, and in Section 4, we present simulations with the hope of quantifying the proportions of our success which were due to skill and chance. Section 5 summarizes and concludes.
 
\section{NCAA tournament modeling}

\subsection{Data selection}

Two easily accessible sets of predictors for NCAA basketball tournament outcomes are information from prior tournaments and results from regular season competition. Regular season data would generally include information like each game's home team, away team, location, and the final score. For tournament games, additional information would include each team's seed (No. 1 to No. 16), region, and the distance from each school's campus to the game location.

The specific viability of using team seed to predict tournament success has been examined extensively; see, for example, \cite{schwertman1996more} and \cite{boulier1999sports}. In place of team seeds, which are approximate categorizations of team strengths based mostly on perceived talent, we supplemented regular season data with two types of information that we thought would be more relevant towards predicting tournament outcomes: (1) the Las Vegas point spread and (2) team efficiency metrics. 

\subsubsection{The Las Vegas point spread}
One pre-game measurement available for the majority of Division 1 men's college basketball games over the last several seasons is the Las Vegas point spread. This number provides the predicted difference in total points scored between the visiting and the home team; a spread of -5.5, for example, implies that the home team is favored to win by 5.5 points. To win a wager placed on a 5.5 point favorite, one would need that squad to win by six points or more. Meanwhile, a bet on the underdog at that same point spread would win either if the underdog loses by 5 points or fewer, thereby covering the spread, or if the underdog outright wins. In principal, the point spread accounts for all pre-game factors which might determine the game's outcome, including relative team strength, injuries, and location. 

Rules of efficient gambling markets imply that, over the long run, it is nearly impossible to outperform the point spreads set by sportsbooks in Las Vegas. A few landmark studies, including \cite{harville1980predictions} and \cite{stern1991probability}, used data from National Football League (NFL) games to argue that, in general, point spreads should act as the standards on which to judge any pre-game predictions. While recent work has looked at gambling markets within, for example, European soccer \citep{constantinou2013profiting}, the Women's National Basketball Association \citep{paul2014market}, the NFL \citep{nichols2014impact}, and NCAA men's football \citep{linna2014effects}, most research into the efficiency of men's college basketball markets was produced several years ago. \cite{colquitt2001testing}, for example, argued that, overall, evidence of market inefficiencies in men's college basketball were limited. These authors also found higher degrees of efficiency in betting markets among contests in which a higher amount of pre-game information was available. \cite{paul2005market} highlighted inefficiencies with respect to larger point spreads using men's college basketball games played between 1996-1997 and 2003-2004, and found that placing wagers on heavy underdogs could be profitable. Lastly, \cite{carlin1996improved} modeled tournament outcomes from the 1994 NCAA season, finding that the point spread was among the easiest and most useful predictors. 

As a result of our relative confidence in the efficiency of men's basketball markets, we extracted the point spread from every Division 1 men's basketball contest since the 2002-2003 season using www.covers.com, and linked these results to a spreadsheet with game results. 

\subsubsection{Efficiency metrics}

One aspect lost in the final scores of basketball games is the concept of a possession. Given that NCAA men's teams have 35 seconds on each possession with which to attempt a shot that hits the rim, the number of possessions for each team in a 40-minute game can vary wildly, depending on how quickly each squad shoots within each 35-second window. In the 2013-2014 season, for example, Northwestern State led all of Division 1 with 79.3 possessions per game, while Miami (Florida) ranked last of the 351 teams with 60.6 per game \citep{TR}. As a result, it is not surprising that Northwestern State scored 20.1 more points per game than Miami, given the large discrepancy in each team's number of opportunities \citep{TR}. As score differentials will also be impacted by the number of possessions in a game, offensive and defensive per-possession scoring rates may provide a greater insight into team strength, relative to the game's final score. 

Several examples of possession-based metrics can be found on a popular blog developed by Ken Pomeroy (www.kenpom.com). Pomeroy provides daily updated rankings of all Division 1 teams, using offensive and defensive efficiency metrics that he adjusts for game location and opponent caliber. The larger umbrella of possession-based statistics, of which Pomeroy's metrics fall under, are summarized by \cite{kubatko2007starting}. 

Pomeroy's website provides team-specific data for all seasons since 2001-2002. We extracted several different variables that we thought would plausibly be associated with a game's results, including a team's overall rating and its possession-based offensive and defensive efficiencies. These metrics provide a unique summary of team strength at each season; one downside, however, is that the numbers that we extracted were calculated $after$ tournament games, meaning that postseason outcomes were included. As a result, fitting postseason outcomes using Pomeroy's end-of-postseason numbers may provide too optimistic a view of how well his numbers produced at the end of the regular season would do. Given that there are many more regular season games than postseason games, however, we anticipated that changes between a team's possession-based efficiency metrics, as judged at the end of the regular season and at the end of the postseason, would be minimal. Relatedly, \cite{kvam2006logistic} found that most of the variability in a team's success during the tournament could be explained by games leading up to mid-February, implying that games at the end of the regular season do not have a dramatic impact on evaluation metrics.

\subsection{Contest requirements}

Standard systems for scoring NCAA basketball tournament pools, including those used in contests hosted online by ESPN (11 million participants in 2014) and Yahoo (15 million), award points based on picking each tournament game winner correctly, where picks are made prior to start of the tournament \citep{espn, yahoo}. In these pools, there are no lost points for incorrect picks, but it is impossible to pick a game correctly if you had previously eliminated both participating teams in earlier rounds. The standard point allocation ranges from 1 point per game to 32 points for picking the tournament winner, or some function thereof, with successive rounds doubling in value. With the final game worth 32 times each first round game, picking the eventual tournament champion is more or less a prerequisite for a top finish. For example, among roughly one million entries in one 2014 ESPN pool, the top 106 finishers each correctly pegged the University of Connecticut as the champion \citep{pagels}.  In terms of measuring the best prognosticator of all tournament games, the classic scoring system is inadequate, leading some to call for an updated structure among the websites hosting these contests \citep{pagels}; for more on optimal strategies in standard pools, see \cite{metrick1996march} and \cite{breiter1997play}. 

Systems that classify games as `win' or `lose' fail to provide probability predictions, and without probabilities, there is no information provided regarding the strength of victory predictions.  For example, a team predicted to win with probability 0.99 by one system and 0.51 by another would both yield a `win' prediction, even though these are substantially different evaluations. An alternative structure would submit a probability of victory for each participating team in each contest. In the Kaggle contest, for example, each participant's submissions consisted of a 2278 x 2 file. The first column contained numerical identifications for each pair of 2014 tournament qualifiers, representing all possible games which could occur in the tournament. For simplicity, we refer to these teams as Team 1 and Team 2. The second column consisted of each submission's estimated probability of Team 1 defeating Team 2. Alphabetically, the first possible match-up in the 2014 tournament pitted the University of Albany (Team 1) against American University  (Team 2); like 2213 of the other possible match-ups, however, this game was never played due to the tournament's single elimination bracket structure. 

There were 433 total entries into the 2014 Kaggle contest, submitted by 248 unique teams. Each team was allowed up to 2 entries, with only the team's best score used in the overall standings. Let $\hat{y}_{ij}$ be the predicted probability of Team 1 beating Team 2 in game $i$ on submission $j$, where $i = 1, ...2278$ and $j = 1, ... , 433$, and let $y_i$ equal 1 if Team 1 beats Team 2  in game $i$ and 0 otherwise.   Each Kaggle submission $j$ was judged using a log-loss function, $LogLoss_{j}$, where, letting $I(Z_i=1)$ be an indicator for whether or not game $i$ was played, 
\begin{eqnarray}
LogLoss_{ij}& = & -\bigg(y_i \log(\hat{y}_{ij}) + (1 - y_i) \log(1 - \hat{y}_{ij})\bigg)*I(Z_i=1) \label{LL} \\ 
LogLoss_{j} & = & \frac{1}{\sum_{i = 1} ^ {2278}  I(Z_i=1)} \sum_{i=1}^{2278} \left[ LogLoss_{ij} \right] \\
& = & \frac{1}{63} \sum_{i=1}^{2278} \left[ LogLoss_{ij} \right].
\end{eqnarray}

\noindent Smaller log-loss scores are better, and the minimum log-loss score (i.e., picking all games correctly with probability 1) is 0. Only the 63 games which were eventually played counted towards the participants' standing; i.e., $\sum_{i = 1} ^ {2278}  I(Z_i=1) = 63$. 

There are several unique aspects of this scoring system. Most importantly, the probabilities that minimize the log-loss function are the same as the probabilities that maximize the likelihood of a logistic regression function. As a result, we begin our prediction modeling by focusing on logistic regression. 

Further, all games are weighted equally, meaning that the tournament's first game counts as much towards the final standings as the championship game. Finally, as each entry picks a probability associated with every possible contest, prior picks do not prevent a submission from scoring points in future games. 

\subsubsection{Predicting NCAA games under probability based scoring function}

Despite the intuitiveness of a probability based scoring system, little research has explored NCAA men's basketball predictions based on the log-loss or related functions. In one example that we build on, \cite{carlin1996improved} supplemented team-based computer ratings with pre-game spread information to improve model performance on the log-loss function in predicting the 1994 college basketball tournament. This algorithm showed a better log-loss score when compared to, among other methods, seed-based regression models and models using computer ratings only. However, Carlin's model was limited to computer ratings based only on the final scores of regular season games, and not on possession-based metrics, which are preferred for basketball analysis \citep{kubatko2007starting}. Further, the application was restricted to the tournament's first four rounds in one season, and may not extrapolate to the final two rounds or to other seasons.

\cite{kvam2006logistic} applied similar principles in using logistic regression as the first step in developing a team ranking system prior to each tournament and found that their rankings outperformed seed-based evaluation systems. However, this proposal was more focused on picking game winners than improving scoring under the log-loss function.

\section{Model selection}
\label{MC}
Our submission was based on two unique sets of probabilities, $\boldsymbol{\hat{y}_{m_1}}=\left[\hat{y}_{1,m_1}, ... \hat{y}_{2278,m_1}\right]$ and $\boldsymbol{\hat{y}_{m_2}}=\left[\hat{y}_{1,m_2}, ... \hat{y}_{2278,m_2}\right]$, generated using a point spread-based model ($M_1$) and an efficiency-based model ($M_2$), respectively.

For $M_1$, we used a logistic regression model with all Division 1 NCAA men's basketball games from the prior 12 seasons for which we had point spread information. For game $g$, $g = 1, ..., 65043$, let $y_g$ be our outcome variable, a binary indicator for whether or not the first team (Team 1) was victorious. Our only covariate in $M_1$ is the game's point spread, $spread_g$, as shown in Equation (\ref{m1}),   
\begin{eqnarray}
\text{logit}(Pr(y_g = 1)) = \beta_0 + \beta_1 * spread_g \label{m1}.  \end{eqnarray}
We used the maximimum likelihood estimates of $\beta_0$ and $\beta_1$ from Equation (\ref{m1}), $\hat{\beta}_{0,m_1}$ and $\hat{\beta}_{1,m_1}$, to calculate $\hat{y}_{i, m_1}$ for any $i$ using $spread_i$, the point spread for 2014 tournament game $i$ such that 
\begin{equation} \hat{y}_{i,m_1} = \hat{\pi}_i = \frac{\text{exp}^{\hat{\beta}_{0,m_1}+\hat{\beta}_{1,m_1}* spread_i}}{1+\text{exp}^{\hat{\beta}_{0,m_1}+\hat{\beta}_{1,m_1}*spread_i}}. \end{equation}
The actual point spread was available for the tournament's first 32 games; for the remaining 2246 contests, we predicted the game's point spread using a linear regression model with 2013-2014 game results.\footnote{This specific aspect of the model has been previously used for proprietary reasons, and we are unfortunately not at liberty to share it.} Of course, just 31 of these 2246 predicted point spreads would eventually be needed, given that there are only 31 contests played in each tournament after the first round. 

An efficiency model ($M_2$) was built using logistic regression on game outcomes, with seven team-based metrics for each of the game's home and away teams as covariates, along with an indicator for whether or not the game was played at a neutral site. These covariates are shown in Table \ref{KP}. Each team's rating represents its expected winning percentage against a league average team \citep{KP}. Offensive efficiency is defined as points scored per 100 possessions, defensive efficiency as points allowed per 100 possessions, and tempo as possessions per minute. Adjusted versions of offensive efficiency, defensive efficiency, and tempo are also shown; these standardize efficiency metrics to account for opposition quality, site of each game, and when each game was played \citep{KP}.  

\begin{table*}
\centering
\caption{Team-based efficiency metrics}
\label{KP}
\begin{tabular}{l l l}
\toprule  
 Variable & Description & Team \\
\hline 
$X_1$&Rating & Home  \\
$X_2$&Rating & Away  \\
$X_3$&Offensive Efficiency & Home \\
$X_4$&Offensive Efficiency & Away \\
$X_5$&Defensive Efficiency & Home  \\
$X_6$&Defensive Efficiency & Away \\
$X_7$&Offensive Efficiency, Adjusted & Home  \\
$X_8$&Offensive Efficiency, Adjusted & Away  \\
$X_9$&Defensive Efficiency, Adjusted & Home \\
$X_{10}$&Defensive Efficiency, Adjusted & Away  \\
$X_{11}$&Tempo & Home  \\
$X_{12}$&Tempo & Away \\
$X_{13}$&Tempo, Adjusted & Home  \\
$X_{14}$&Tempo, Adjusted & Away  \\
$X_{15}$&Neutral & N/A \\
\bottomrule
\end{tabular}
\end{table*}

We considered several different logistic regression models, using different combinations and functions of the 15 variables in Table \ref{KP}. Our training data set, on which models were fit and initial parameters were estimated, consisted of every regular season game held before March 1, using each of the 2002-2003 through 2012-2013 seasons. For our test data, on which we averaged the log-loss function in Equation (\ref{LL}) and selected our variables for $M_2$, we used all contests, both regular season and postseason, played after March 1 in each of these respective regular seasons. We avoided only using the Division 1 tournament outcomes as test data because only about 1\% of a season's contests are played during these postseason games. Given that March includes conference tournament games, which are perhaps similar to those in the Division 1 tournament, and our desire to increase the pool of test data, we chose the earlier cutoff. Table \ref{Modelbuild} shows examples of the models we considered and their $LogLoss$ score averaged on the test data. 

\begin{table*}
\centering
\caption{Model building results}
\label{Modelbuild}
\begin{tabular}{l l l}
\toprule  
 Fit & Variables & $LogLoss$\^{}\\
\midrule 
(a)&($X_1$ - $X_2$) & 0.509\\
(b)&($X_1$ - $X_2$), $X_{15}$ & 0.496\\
(c)&$X_1$, $X_2$ & 0.510\\
(d)&$X_1$, $X_2$, $X_{15}$ & 0.496\\
(e)&$X_3$, $X_4$, $X_5$, $X_6$, $X_{15}$ & 0.538\\
\hlc[green]{(f)}&$X_7$, $X_8$, $X_9$, $X_{10}$, $X_{15}$  &\hlc[green]{0.487}\\
(g)&($X_7$ - $X_8$), ($X_9$ - $X_{10}$), $X_{15}$  &0.487\\
(h)&$X_1$, $X_2$, $X_7$, $X_8$, $X_9$, $X_{10}$, $X_{15}$  & 0.487\\
(i)**&($X_7$, $X_8$, $X_9$, $X_{10}$, $X_{15})^2$  & 0.488\\
(j)**&($X_1$, $X_2$, $X_7$, $X_8$, $X_9$, $X_{10}$, $X_{13}$, $X_{14}$ , $X_{15})^2 $ & 0.488\\
(k)***&($X_1$, $X_2$, $X_7$, $X_8$, $X_9$, $X_{10}$, $X_{13}$, $X_{14}$ , $X_{15})^3 $ & 0.493\\
\hline
 \multicolumn{3}{l}{\^{}\ Games after March 1, in each of the 2002-2003 to 2012-2013 seasons}  \\
 \multicolumn{3}{l}{**\ all two-way interactions of these variables}  \\
 \multicolumn{3}{l}{***\ all three-way interactions of these variables}  \\ 
 \multicolumn{3}{l}{Chosen model is \hlc[green]{highlighted}}  \\ \bottomrule
\end{tabular}
\end{table*}

While not the complete set of the fits that we considered, Table \ref{Modelbuild} gives an accurate portrayal of how we determined which variables to include. First, given the improvement in the loss score from fits (a) to (b) and (c) to (d), inclusion of $X_{15}$, an indicator for if the game was played on a neutral court, seemed automatic. Next, fit (f), which included the overall team metrics that had been adjusted for opponent quality, provided a marked improvement over the unadjusted team metrics in fit (e). Meanwhile, inclusion of overall team rating (fit (h)), and linear functions of team efficiency metrics (fit (g)), failed to improve upon the log-loss score from fit (f). Higher order terms, as featured in models (i), (j), and (k), resulted in worse log-loss performances on the test data, and an ad-hoc approach using trial and error determined that there were no interaction terms worth including.

The final model for $M_2$ contained the parameter estimates from a logistic regression fit of game outcomes on $X_7$, $X_8$, $X_9$, $X_{10}$, and $X_{15}$ (Adjusted offensive efficiency for home and away teams, adjusted defensive efficiency for home and away teams, and a neutral site indicator, respectively). We estimated $\boldsymbol{\hat{y}_{m_2}}$ using the corresponding team specific metrics from kenpom.com, taken immediately prior to the start of the 2014 tournament. 

Our final step used ensembling, in which individually produced classifiers are merged via a weighted average \citep{opitz1999popular}. Previous work has shown that ensemble methods work best using accurate classifiers which make errors in different input regions, because areas where one classifier struggles would be offset by other classifiers \citep{hansen1990neural}. While our two college basketball classifiers, $M_1$ and $M_2$, likely favor some of the same teams, each one is produced using unique information, and it seems plausible that each model would offset areas in which the other one struggles. 

A preferred ensemble method takes the additional step of calculating the optimal weights \citep{dietterich2000ensemble}. Our chosen weights were based on evidence that efficiency metrics were slightly more predictive of tournament outcomes than the model based on spreads. Specifically, using a weighted average of $M_1$ and $M_2$,  we calculated a $LogLoss$ score averaged over each of the Division 1 tournaments between 2008 and 2013 (incidentally, this was the `pre-test' portion of the Kaggle contest). The balance yielding the best $LogLoss$ score gave a weight of 0.69 to $M_2$ and 0.31 to $M_1$.

Thus, we wanted one of our submissions to give more importance to the efficiency model. However, given that each season's efficiency metrics may be biased because they are calculated after the tournament has concluded, for our other entry, we reversed the weightings to generate our two submissions, $\boldsymbol{S_1}$ and $\boldsymbol{S_2}$, rounding our weights for simplification. 
\begin{eqnarray}	\boldsymbol{S_1} = 0.75*\boldsymbol{\hat{y}_{m_1}} + 0.25* \boldsymbol{\hat{y}_{m_2}} \nonumber \label{WTE}\\
	\boldsymbol{S_2} = 0.25*\boldsymbol{\hat{y}_{m_1}} + 0.75* \boldsymbol{\hat{y}_{m_2}} \nonumber \label{WTE2}\end{eqnarray}

The correlation between $\boldsymbol{S_1}$ and $\boldsymbol{S_2}$ was 0.94, and 78\% of game predictions on the two entries were within 0.10 of one another. Our top submission, $\boldsymbol{S_2}$, finished in first place in the 2014 Kaggle contest with a score of 0.52951. Submission $\boldsymbol{S_1}$, while not officially shown in the standings as it was our second best entry, would have been good enough for fourth place (score of 0.54107).

\section{Simulation Study}

In order to evaluate the luck involved in winning a tournament pool with probability entries, we performed a simulation study, assigning each entry a $LogLoss$ score at many different realizations of the 2014 NCAA basketball tournament. The contest organizer provided each of the 433 submissions to the 2014 Kaggle contest for this evaluation. 

To simulate the tournament, ``true" win probabilities must be specified for each game. We evaluate tournament outcomes over five sets of true underlying game probabilities: $\boldsymbol{S_1}$, $\boldsymbol{S_2}$, M($\boldsymbol{S_{All}})$, M($\boldsymbol{S_{Top10}})$, and $\boldsymbol{0.5}$, listed as follows.

\begin{itemize}
\item Our first entry ($\boldsymbol{S_1}$)
\item Our second entry ($\boldsymbol{S_2}$)
\item Median of all Kaggle entries (M($\boldsymbol{S_{All}})$)
\item Median of the top 10 Kaggle entries (M($\boldsymbol{S_{Top10}})$)
\item All games were a coin flip (i.e. $p=0.5$ for all games) ($\boldsymbol{0.5}$)
\end{itemize}

Let rank($\boldsymbol{S_1}$) and rank($\boldsymbol{S_2}$) be vectors containing the ranks of each of our submissions across the 10,000 simulations at a given set of game probabilities. We are interested in the median rank and percentiles (2.5, 97.5) for each submission (abbreviated as M (2.5 - 97.5)), across all simulations. We are also interested in how often each submission finishes first and in the top 10. 

\begin{landscape}
\begin{table*}
\centering
\caption{Simulation results}
\label{Tab2}
\begin{tabular}{l l l l l l l}
\toprule  
 & & \multicolumn{5}{c}{True Game Probabilities} \\ \cline{3-7}
Outcome   & Type & $\boldsymbol{S_1}$ & $\boldsymbol{S_2}$  & M($\boldsymbol{S_{Top10}})$ &M($\boldsymbol{S_{All}})$ & $\boldsymbol{0.5}$\\
\hline 
rank($\boldsymbol{S_1}$)& M (2.5 - 97.5) & 11 (1-168) & 59 (1-202) & 99 (2-236)  &  145 (4-261)  &  264 (186-299) \\
rank($\boldsymbol{S_2}$)& M (2.5 - 97.5)& 53 (2-205) &  14 (1-164)& 92 (2-245) & 146 (5-266)  & 226 (135-285) \\
rank($\boldsymbol{S_1}$) = 1& \%&    15.57 &3.90 & 2.02 &  0.88 & 0 \\
rank($\boldsymbol{S_2}$) = 1& \%&   2.22 &11.65 & 1.89 & 0.63 & 0 \\
rank($\boldsymbol{S_1}) \leq  10$& \%& 48.79 &  17.69& 8.85   &5.04 & 0 \\
rank($\boldsymbol{S_2}) \leq  10$& \%&20.72 &  44.47& 11.96 & 4.77& $<$0.01 \\
Unique winners  & Total & 332 & 336 & 337 &348 &   217\\
\hline 
\multicolumn{7}{l}{M: Median, 2.5: 2.5th percentile, 97.5: 97.5th percentile}\\ \bottomrule
\end{tabular}
\end{table*}
\end{landscape}
\normalsize

Lastly, we extract the number of unique winners across the simulations, which can give us a sense of how many entries had a reasonable chance of winning at each set of underlying game probabilities. 

The results of the simulations appear in Table \ref{Tab2}.  Each column represents a different ``true" probability scenario and each row records the results of a statistic of interest.  The first and second rows show results of our first and second entry, respectively.  We can see that if the ``true" probabilities were $\boldsymbol{S_1}$, our entry finished at a median of 11th place, whereas if the true probabilities were $\boldsymbol{S_2}$, our median finish was 14th place. If the true probabilities were $\boldsymbol{S_1}$, our entry containing those probabilities would finish in first place around 15\% of the time.  Likewise, with $\boldsymbol{S_2}$ as ``true" probabilities, that entry would win around 12\% of the time.  Relative to a contest based entirely on luck, where each entry would have a 1 in 433 chance of finishing first, our chances of winning were roughly 50 to 60 times higher using $\boldsymbol{S_1}$ and $\boldsymbol{S_2}$ as the truth. This conceivably represents the upper bound of our submission's `skill.'

On the whole, our simulations indicated that the amount of luck required to win a contest like the Kaggle one is enormous; even if you knew the true probabilities of a win for every single game with certainty, you'd still only win about 1 in 8 times! In fact, even if our submissions were  correct, we'd only finish in the top 10 about 49\% and 45\% of the time, respectively, for $\boldsymbol{S_1}$ and $\boldsymbol{S_2}$.

If the median of all entries (M($\boldsymbol{S_{All}})$) or the median of the top 10 entries (M($\boldsymbol{S_{Top10}})$) is used as the true probabilities, our chances of winning diminish.  For M($\boldsymbol{S_{Top10}})$, our chances of winning on entries $\boldsymbol{S_1}$ and $\boldsymbol{S_2}$ were both about 2\%.  For M($\boldsymbol{S_{All}})$, our chances of winning on entries $\boldsymbol{S_1}$ and $\boldsymbol{S_2}$ were both less than 1\%.  Lastly, if each game was truly a coin flip, neither of our entries finished first in any of the simulations. 

Of the 433 total entries, fewer than 350 finished in first place at least once in each of the simulations with our entries as the truth.  This suggests that if either of our submitted probabilities were close to the ``true'' probabilities, about 20\% of entries had little to no chance of winning.

Lastly, Figure 1 shows the smoothed density estimates (dark line) of all winning $LogLoss$ scores from 10,000 simulated tournaments under the game probabilities in $\boldsymbol{S_2}$, along with the density estimates for $\boldsymbol{S_2}$ on simulations in which that entry was the winner. The winning score of $\boldsymbol{S_2}$ in 2014 (0.529) is shown by a vertical line. Relative to the simulated winning scores, $\boldsymbol{S_2}$ winning scores have a lower density in the tails.  The 2014 winning score was relatively higher than most of the scores that won the simulated tournaments, perhaps because the University of Connecticut, a seven seed, won all six of its games en route to becoming Division 1 champions. Prior to Connecticut's win, only four previous champions since 1979 were seeded lower than three (four seed Arizona (1997), six seeds Kansas (1988) and North Carolina (1983), and eight seed Villanova (1985)).  As a result, most predictive models would have a seven seed as a substantial underdog in games against higher seeded opponents in the later rounds, leading to comparatively larger values of the loss function than when favorites prevail.

\begin{figure}\begin{center}
***********************Insert Figure 1 here***********************\end{center}
 \caption{Smoothed density estimates of winning scores across 10,000 tournament simulations using underlying game probabilities $\boldsymbol{S_2}$. The dotted line refers to the winning scores on simulations won by the $\boldsymbol{S_2}$ entry.}\end{figure}

\section{Conclusion}
While traditional NCAA basketball tournament bracket pools are here to stay, Kaggle has developed an alternative scoring system that requires a probability prediction rather than simply picking a winner. Given these guidelines, we used Las Vegas point spread data and Ken Pomeroy's efficiency ratings to build predictive models that ultimately led to a first place finish in this contest.  

We employed logistic regression models, based, in part, on the fact that the maximum likelihood estimates derived from logistic regression are based on maximizing a function that was equivalent to the contest scoring function.  While logistic regression is a fairly standard statistical technique, we propose that it was important in this context specifically because of the scoring function.  

While our choice of a class of models possibly played a role in our victory, our choice of data likely played a much larger role.  It is extremely difficult to generate predictive models that outperform the Las Vegas point spread, particularly in high profile games like the ones in the NCAA tournament, and both the point spread and efficiency ratings have previously been shown to work well in predicting college basketball outcomes \citep{carlin1996improved}. Conceptually, one could argue that the Las Vegas point spread is a subjective prior based on expert knowledge, whereas Pomeroy's ratings are based entirely on data.  In this way, our ensembling of these two sources of data follows the same principles as a Bayesian analysis.  

Given the size of the Kaggle contest, it is reasonable to estimate that our models increased our chances of winning anywhere from between five-fold to fifty-fold, relative to a contest that just randomly picked a winner. However, even with a good choice of models and useful data, we demonstrated that luck also played a substantial role.  Even in the best scenario where we assumed that one of our predicted probabilities was correct, we found that this entry had less than a 50\% chance of finishing in the top ten and well less than a 20\% chance of winning, given a contest size of 433 entrants. Under different, but fairly realistic true probability scenarios, our chances of winning decreased to be less than 2\%.

\bibliographystyle{DeGruyter}
\bibliography{kaggle}

\end{document}